\documentclass[sigconf]{acmart}
\usepackage{algorithm}
\usepackage{algpseudocode} 
\usepackage{graphics}
\acmConference[SBES '25]{Brazilian Symposium on Software Engineering}{September 22-26, 2025}{Recife, PE}
\AtBeginDocument{%
  }

\setcopyright{none} 
\renewcommand\footnotetextcopyrightpermission[1]{} 
\begin{document}
\settopmatter{printacmref=false}

\title{The Impact of Generative AI on Code Expertise Models: An Exploratory Study}

\author{Otávio Cury}
\email{otaviocury@ufpi.edu.br}
\affiliation{%
  \institution{Federal University of Piauí}
  \city{Teresina}
  \state{Piauí}
  \country{Brazil}
}

\author{Guilherme Avelino}
\email{gaa@ufpi.edu.br}
\affiliation{%
  \institution{Federal University of Piauí}
  \city{Teresina}
  \state{Piauí}
  \country{Brazil}
}
 
\begin{abstract}
Generative Artificial Intelligence (GenAI) tools for source code generation have significantly boosted productivity in software development. However, they also raise concerns, particularly the risk that developers may rely heavily on these tools, reducing their understanding of the generated code. We hypothesize that this loss of understanding may be reflected in source code knowledge models, which are used to identify developer expertise. In this work, we present an exploratory analysis of how a knowledge model and a Truck Factor algorithm built upon it can be affected by GenAI usage. To investigate this, we collected statistical data on the integration of ChatGPT-generated code into GitHub projects and simulated various scenarios by adjusting the degree of GenAI contribution. Our findings reveal that most scenarios led to measurable impacts, indicating the sensitivity of current expertise metrics. This suggests that as GenAI becomes more integrated into development workflows, the reliability of such metrics may decrease.
\end{abstract}

\keywords{generative artificial intelligence, source code expertise, truck factor, mining software repository}

\maketitle

\section{Introduction}

In recent years, Generative Artificial Intelligence (GenAI) has attracted considerable attention from the Software Engineering community, revealing research gaps across multiple areas \cite{nguyen2023generative}. This growing interest is justified by GenAI's potential to transform a range of domains \cite{gozalo2023survey}. In software development, the increasing integration of GenAI into development environments has brought numerous benefits. Tools such as OpenAI's ChatGPT\footnote{https://chatgpt.com} and GitHub Copilot\footnote{https://github.com/features/copilot} have contributed to increased productivity and enhanced software quality \cite{dohmke2023sea, ziegler2024measuring}. Despite these advantages, researchers have raised concerns related to adopting such technologies \cite{ernst2022ai}.

One significant drawback is the risk of developer over-reliance on GenAI tools, where users may accept generated code without fully understanding it \cite{zhang2023demystifying, yilmaz2023augmented}. Such behavior can undermine problem-solving skills and hinder a deeper comprehension of the codebase \cite{prather2023s}. Given this, we argue that \textit{models} used to estimate developer expertise should reflect the potential impact of GenAI-assisted code generation on code understanding. Typically based on authorship data of \textit{Version Control Systems}, these models estimate a developer’s knowledge of the code and are applied across various stages of the software development \cite{hannebauer2016automatically, cury2022identifying}. They also play a key role in knowledge loss mitigation, being used in metrics such as the \textit{Truck Factor}, which estimates the concentration of knowledge within a development team, a well-established line of research \cite{cury2024source, avelino2015truck, jabrayilzade2022bus}.

Building on this premise, this paper investigates how developers' use of GenAI tools may impact source code expertise identification models. For this investigation, we searched for ChatGPT shared links embedded in source code files from GitHub projects. Using these links, we retrieved the corresponding conversations and identified the extent to which GenAI-generated code had been integrated into the project files. Based on statistical evidence of this integration, we simulated different scenarios to assess the potential impact on expertise identification models and a widely used Truck Factor algorithm, by attributing a portion of developer authorship to GenAI. This study is guided by the following overarching question: \textbf{How does the use of GenAI tools for source code generation affect the accuracy and reliability of models that identify developer expertise in source code?}

The key contributions of this study are as follows: \textit{1)} Quantitative insights into how code generated by ChatGPT is integrated into open-source projects; \textit{2)} A comprehensive analysis of how this integration may impact expertise models and their applications, such as knowledge concentration metrics; \textit{3)} A publicly available dataset containing information on the integration of GenAI-generated code in open-source repositories. This work is organized as follows: Section \ref{sec:background} introduces key concepts. Section \ref{sec:related_works} reviews related studies on the relationship between GenAI tools and source code knowledge. Section \ref{sec:study_design} describes the data collection process used in this study. Section \ref{sec:results} presents the results, Section \ref{sec:discussion} discusses them, and Section \ref{sec:conclusion} concludes the paper.

\section{Background} \label{sec:background}

\subsection{Code Knowledge Models} \label{sec_code_knwoledge}

Source code knowledge models are designed to identify developer expertise within software projects. This information supports a range of activities, including task assignment and bug fixing, and also assists project managers in monitoring knowledge concentration within the codebase \cite{ferreira2017comparison, robillard2021turnover}. Most existing approaches rely on development history stored in \textit{Version Control Systems}. For example, the \textit{Degree of Knowledge} model, proposed by Fritz et al., combines two types of information: the developer’s authorship of a file captured by the \textit{Degree of Authorship} (DOA), and the number of interactions the developer has had with that file \cite{fritz2014degree}.

\subsubsection{Degree of Expertise} \label{sec_doe}

In a previous study, we proposed the \textit{Degree of Expertise} (DOE), a knowledge model that uses four variables from development history to measure a developer’s knowledge on a source code file \cite{cury2022identifying}. Unlike existing models in the literature, \textit{DOE} combines fine-grained measures of change, authorship, recency of modification, and file size to achieve greater precision in knowledge estimation. The \textit{DOE} model showed superior performance in identifying file experts \cite{cury2022identifying}, and was applied within a Truck Factor algorithm \cite{cury2024source, cury2024knowledge}. The \textit{DOE} of a developer \textit{d} in version \textit{v} of file \textit{f} is calculated using Equation \ref{eq:doe}.

\begin{align}
\textbf{DOE(d, f(v))} = & \; 5.28223 + 0.23173 \cdot \ln(1 + \textbf{Adds}^{d, f(v)}) \nonumber \\
& + 0.36151 \cdot (\textbf{FA}^{f}) \nonumber - 0.28761 \cdot \ln(\textbf{Size}^{f(v)}) \\
& - 0.19421 \cdot \ln(1 + \textbf{NumDays}^{d, f(v)}) \label{eq:doe}
\end{align}

Where, \textbf{Adds}: number of lines added by developers \textit{d} on file \textit{f}; \textbf{FA}: boolean if developer \textit{d} is the creator of the file \textit{f}; \textbf{Size}: number of lines of code (LOC) of the file \textit{f}; \textbf{NumDays}: number of days since the last commit
of a developer \textit{d} on file \textit{f}. In this work, we examine how the integration of GenAI-generated code impacts this knowledge model. The rationale behind selecting this particular model is further detailed on Section \ref{sec:study_design}. 

\subsection{Truck Factor Algorithm} \label{sec_avl}

The \textit{Truck Factor} measures how many developers would need to leave a software project for it to be critically affected \cite{jabrayilzade2022bus}. This metric helps identify knowledge concentration risks and has been widely studied in the literature \cite{jabrayilzade2022bus, ferreira2017comparison, avelino2016novel, cury2024source}. Among the proposed methods for estimating Truck Factor, Avelino's algorithm \cite{avelino2016novel} is notable for its strong performance in comparison studies \cite{ferreira2017comparison} and for being adopted in subsequent validation studies \cite{calefato2022will, almarimi2021csdetector}.

Avelino’s algorithm estimates the Truck Factor using a strategy based on developer authorship. It identifies file experts through the \textit{Degree of Authorship} (DOA) model and iteratively removes the developer who is the expert for the largest number of files. After each removal, the algorithm checks how many files are left without any expert. This process continues until more than half of the project's files are considered abandoned. The number of developers removed at that point is returned as the \textit{Truck Factor}. The pseudo-code for this algorithm is provided in the original study \cite{avelino2016novel}.

In this study, we use the implementation provided by previous work \cite{cury2024knowledge, cury2024source}, which modifies the original Truck Factor algorithm by replacing the \textit{DOA} model with the \textit{DOE} model (Section~\ref{sec_doe}) for expert identification. This adaptation enables us to investigate how the influence of GenAI-generated code on knowledge models affects knowledge concentration metrics, such as the Truck Factor.

\section{Related Works} \label{sec:related_works}

Some researchers have raised concerns about using Generative Artificial Intelligence (GenAI) to generate source code that might impact programming knowledge. Denny et. al also cite \textit{learner over-reliance}, trusting in the generated code without fully understanding it—as a key risk in student learning \cite{denny2024computing}.

Students also share this concern. Yilmaz et al. analyzed ChatGPT’s limitations from the perspective of undergraduates and found that one common concern was its potential to impair algorithmic thinking skills \cite{yilmaz2023augmented}. Similarly, Ma et al. reported that students in a Python course feared ChatGPT could hinder their learning process \cite{ma2024enhancing}. Prather et al. interviewed students using Copilot in a programming class, and several participants expressed concern that relying on the tool might prevent them from fully understanding the code \cite{prather2023s}. Among practitioners, similar concerns were observed. Russo reported that some engineers fear over-reliance on GenAI tools may reduce comprehension of the code they integrate \cite{russo2024navigating}.

Some studies have employed methodologies similar to ours. Grewal et al. analyzed ChatGPT shared conversation links to investigate how generated code is integrated into software projects \cite{grewal2024analyzing}. Using the \textit{Levenshtein Distance}, they measured the degree to which code from these conversations was incorporated into GitHub repositories. Similarly, Jin et al. examined the effectiveness of ChatGPT in assisting developers with code generation \cite{jin2024can}. Applying a comparable approach, they found that in 16.8\% of conversations, the generated code snippets had exact matches in the main branches of the analyzed projects, allowing for minor modifications.
 
Although we did not identify studies addressing the relationship between GenAI and \textit{Knowledge Models}, the existing literature highlights a research gap. It suggests that using GenAI tools for source code generation may lead to developers not fully comprehending the code generated and integrated into their projects. 

\section{Study Design} \label{sec:study_design}

\begin{figure*}[h]
  \centering
  \includegraphics[width=0.9\linewidth]{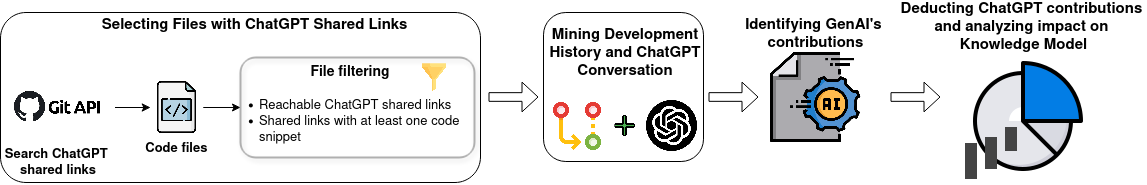}
  \caption{Overview of the methodology used to assess the impact of GenAI on a knowledge model.}
  \label{fig:study_design}
\end{figure*}

Our study design consists of four main steps. First, we select source code files that contain ChatGPT links. Then, we mine the development history of these files and retrieve the code generated in the associated conversations with ChatGPT. By combining these two sources of information, we identify the extent to which code generated by Generative Artificial Intelligence (GenAI) was integrated into the files. Finally, we statistically analyze these integrations. Based on these insights, we simulate different scenarios to assess the impact of GenAI on knowledge models by attributing a portion of the authorship to GenAI rather than to developers. Figure~\ref{fig:study_design} provides an overview of the study design.

\subsection{Selecting Files with ChatGPT Shared Links} \label{sec:select_files_shared_links}

In May 2023, OpenAI introduced a feature that allows users to share their conversations with ChatGPT via shareable links\footnote{https://help.openai.com/en/articles/7925741-chatgpt-shared-links-faq}. This feature enables developers to share their chats with ChatGPT, containing specific solutions integrated into the source code. These shared links appear in GitHub artifacts, and they have already been extracted in related studies \cite{xiao2024devgpt, grewal2024analyzing, jin2024can, hao2024empirical}. Inspired by the study of Xiao et. al \cite{xiao2024devgpt}, in this study, we construct a dataset focused on shared links within GitHub source code files. We chose the \textit{source code files} artifact because it is more closely related to the knowledge metrics used, facilitating the analysis. 

For our initial search, we used the GitHub REST API\footnote{https://docs.github.com/en/rest?apiVersion=2022-11-28} to look for keywords that indicate the presence of ChatGPT shared links within the contents of the source code file. The following endpoint was used to locate these chat links in the source code: \textit{\url{https://api.github.com/search/code?q=\#chatgpt_url+language:\#language}}. We searched for links using two keywords (\#chatgpt\_url). The first, \textit{https://chat.openai.com/share/}, was originally introduced by OpenAI \cite{xiao2024devgpt}. The second, \textit{https://chatgpt.com/share/}, emerged in 2024. To focus on source code files, we applied a programming language filter (\#language), targeting the 10 most popular languages on GitHub in 2022\footnote{https://octoverse.github.com/2022/top-programming-languages}: JavaScript, Python, Java, TypeScript, C\#, C++, PHP, Shell, C, and Ruby. This search, conducted in November 2024, identified 2579 files with shared links. Due to the limitations of the API, which do not allow regex searches, we manually filtered the results to identify artifacts containing links that exactly match the shared link pattern using the following regular expressions: \textit{\url{^https://chat.openai.com/share/[a-zA-Z0-9-]{36}}\$}, and \textit{\url{^https://chatgpt.com/share/[a-zA-Z0-9-]{36}}\$}. After applying these to the file contents, we refined our dataset to 2502 source code files from 1354 repositories, totaling 2036 shared links.

Following the initial search and regex filtering, we applied two additional filters. First, since shared links can be disabled after being created\footnote{https://help.openai.com/en/articles/7925741-chatgpt-shared-links-faq}, we checked the reachability of each valid shared link by fetching the page content and checking for errors. Second, we verified that each successful fetch contained at least one code snippet from ChatGPT. We found 198 shared links without code snippets and 172 disabled links. As a result, our dataset was reduced to 2094 files in 1135 repositories, containing 1666 shared links. The number of source code files and shared links per programming language before and after applying these filters is shown in Table \ref{tab:shared_links_initial_search}. 

\begin{table}[h]
\caption{Number of source code files and shared links per programming language before and after applying the filters.}
\centering
\label{tab:shared_links_initial_search}
\begin{tabular}{@{}lcccc@{}}
\toprule
\multicolumn{1}{c}{}                  & \multicolumn{2}{c}{\textbf{Before Filtering}} & \multicolumn{2}{c}{\textbf{After Filtering}} \\ \midrule
\multicolumn{1}{c}{\textbf{Language}} & \textbf{Files}     & \textbf{Shared Links}    & \textbf{Files}    & \textbf{Shared Links}    \\ \midrule
Python                                & 988                & 803                      & 781               & 629                      \\
JavaScript                            & 542                & 483                      & 476               & 414                      \\
C++                                   & 251                & 195                      & 223               & 169                      \\
Java                                  & 247                & 183                      & 210               & 146                      \\
TypeScript                            & 178                & 134                      & 146               & 108                      \\
Shell                                 & 100                & 90                       & 88                & 79                       \\
C\#                                   & 80                 & 62                       & 69                & 51                       \\
C                                     & 68                 & 63                       & 59                & 53                       \\
PHP                                   & 45                 & 27                       & 41                & 22                       \\
Ruby                                  & 3                  & 4                        & 1                 & 1                        \\ \bottomrule
\end{tabular}
\end{table}

The results show a significantly higher number of links in Python files, 60\% more than in the second most common language, JavaScript. This finding aligns with Xiao \cite{xiao2024devgpt}. Additionally, Python and JavaScript are among the languages whose users most frequently utilize ChatGPT and Copilot \cite{peslak2024ai, ziegler2024measuring}.

After applying these filters, the distribution of the number of files with shared links per repository has a first quartile (Q1) of 1, a median (Q2) of 1, and a third quartile (Q3) of 1. Similarly, the shared link distribution per repository has Q1 = 1, Q2 = 1, and Q3 = 2. These statistics indicate sparse distributions, where most repositories contain only one file with shared links, and most of these files contain only a single shared link.

\subsection{Extracting Development History and GenAI Conversations}

After filtering all relevant links and associated files, we constructed a dataset to analyze how the solutions provided in the links were integrated into their respective files. This step allowed us to assess the extent to which the generated code was adopted and how much of the developers’ contributions could be attributed to ChatGPT.

Using the successfully fetched shared links described in Section \ref{sec:select_files_shared_links}, we first extracted the developers’ prompts with the corresponding ChatGPT responses that included generated code snippets. We then cloned 1,135 repositories to collect the development history of the files containing the shared links. The vast majority of these repositories were relatively small, with a median of 1 contributor (Q1 = 1, Q3 = 2), 24 commits, and 34 files.

\subsection{Identifying GenAI's contributions}

In addition to mining data to calculate the author's contributions to the files that contain the links, we also needed to assess how much of the code from each link was integrated into the file. Following the study by Grewal et al. \cite{grewal2024analyzing}, we identified \textit{matched lines} between code snippets and lines added in the same commit as the shared link. While their study used \textit{Levenshtein Distance} for fuzzy matching, we take a more conservative approach and consider \textit{only exact matches}. Our analysis is based on two assumptions, explained below with their supporting rationale.
\\

\noindent\fbox{%
    \parbox{0.46\textwidth}{%
        \textit{\textbf{Assumption 1:} The exact matching lines identified correspond to instances of code directly copied and pasted from ChatGPT-generated outputs.}
    }%
}\\

\noindent \textbf{Rationale:} Since there is no known direct integration of ChatGPT with development environments, we consider lines that match the generated code as instances of \textit{copying}. In the context of Copilot, matched lines are often referred to in research as \textit{accepted} lines. A previous study by Dohmke found that developer satisfaction and productivity increase as the acceptance rate rises \cite{dohmke2023sea}. With the growing use of ChatGPT for code generation \cite{stackOverflow2024, brachman2025current}, we can assume similar trends, although this may reduce developers’ autonomy over the code \cite{bird2023taking}. Moreover, research on Copilot has already shown an increase in copied and repeated code \cite{harding2025aicopilot}, while studies on ChatGPT indicate that more than 50\% of generated code snippets are integrated without modifications \cite{grewal2024analyzing}.
\\

\noindent\fbox{%
    \parbox{0.46\textwidth}{%
        \textit{\textbf{Assumption 2:} Developers do not acquire the same level of knowledge from integrating copied ChatGPT-generated code.}
    }%
}\\

\noindent \textbf{Rationale:} Passively accepting solutions may impact the understanding of integrated code \cite{prather2023s} and create cognitive dissonance regarding one's problem-solving abilities, leading to an illusion of competence, as demonstrated by Prather's study with students \cite{prather2024widening}. We believe this effect should be reflected in knowledge models. Research suggests that actively retyping a solution enhances comprehension and learning \cite{gaweda2020typing, skripchuk2023analysis}. For example, active code retyping has been a key component of proposed methodologies for improving cognitive engagement with AI-generated code \cite{kazemitabaar2024exploring}.

Building on these assumptions, we quantify the extent of copied code for each \textit{file–shared link} pair. This provides statistical evidence of how ChatGPT-generated code is integrated into open-source projects and allows us to simulate different usage scenarios by attributing varying levels of authorship to GenAI.

\subsection{Analyzing GenAI Conversations and Code Integration} \label{sec:shared_link_analysis}

Among the 2,235 file–shared link pairs selected and filtered (Section \ref{sec:select_files_shared_links}), 1,699 (76\%) contained at least one matched line. However, manual inspection revealed that some of these matches consisted only of single-character overlaps, such as individual symbols or keys. To improve data quality, we applied an additional filter to retain only pairs with at least one matched line containing more than one character. After this refinement, 1,672 pairs remained (74\%), and only these will be considered in the subsequent analysis.

With this filter applied, we primarily analyzed the distribution of the percentage of code copied from ChatGPT by developers and how this varies across programming languages. Additionally, we assessed the number of conversation turns present in these solutions, a characteristic frequently explored in related studies. 

\subsection{Impact Simulation Design} \label{sec:impact_measurement_design}

As demonstrated in the previous sections, the dataset we constructed is sparse regarding the link–file–repository relationship. In most cases, a single link is added in a single commit to a single file within a repository. As a result, a direct analysis of the impact of copied code from these links on files and repository history would be limited and might not reflect realistic usage scenarios. Therefore, drawing on the statistical findings from the real data (presented in Section \ref{sec:code_integration_statistics}), we performed a \textit{simulated} analysis to assess the potential impact of copied code.

Instead of simulating the impact on the repositories containing shared links, most of which are small and have limited relevance, we focused on more prominent repositories. Specifically, we selected the five most-starred GitHub projects for each of the ten chosen programming languages. We excluded 14 non-software repositories (e.g., code collections and roadmaps) due to their limited relevance for knowledge concentration analysis, and 12 additional projects were excluded based on size, following methodologies from related studies \cite{avelino2016novel, cury2022identifying}. Ultimately, 24 repositories were selected, as shown in Table \ref{tab:target_repositories_impact} (Section \ref{sec:results}).

We simulated the impact by applying a \textit{uniform code copy rate}, derived from the statistical analysis presented in Section \ref{sec:code_integration_statistics}. For this simulation, we adopted the Degree of Expertise (DOE) model for two primary reasons. First, \textit{DOE} estimates a developer’s expertise in a file partially based on the number of lines they have added, which enables us to discount lines attributed to GenAI. This level of granularity is not supported by models like Degree of Authorship (DOA), which rely on commit counts. Second, previous research has demonstrated that \textit{DOE} outperforms other models in identifying developer expertise and in computing the Truck Factor of software projects \cite{cury2024source}.

We analyzed the impact of code copying on two key aspects within the scope of this work. First, we investigated how code copying affects the \textit{DOE} values by evaluating the impact of excluding these contributions on the expertise scores of developers. Second, we examined how these changes influence the Truck Factor of the projects, analyzing alterations in the ranking of Truck Factor developers and changes in the Truck Factor values themselves.

\begin{figure}[h]
    \centering
    \includegraphics[width=\linewidth]{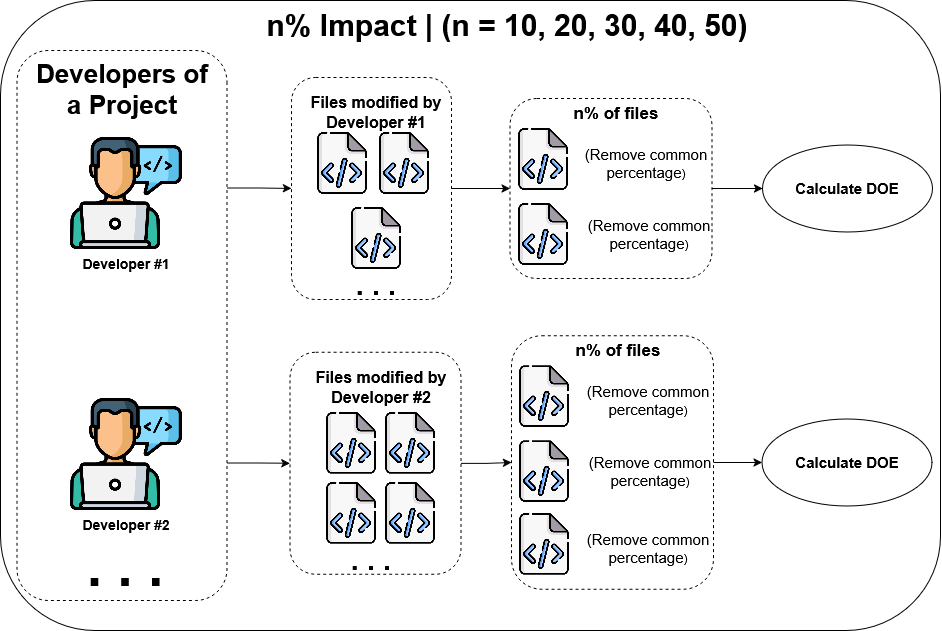}
    \caption{Overview of the approach used to simulate the impact of GenAI code copying across different usage scenarios.}
    \label{fig:gen_ai_impact_analysis}
\end{figure}

For the Truck Factor analysis, we evaluated the impact by varying the percentage of files affected by code copying to 10\%, 20\%, 30\%, 40\%, and 50\%. In each iteration, the affected files for each developer were randomly selected. Then, in every commit made by that developer on these files, we deducted the uniform code copy rate of the added lines, simulating GenAI authorship. Figure \ref{fig:gen_ai_impact_analysis} illustrates the procedure. In each iteration, the Truck Factor was compared to the original results without GenAI impact. For the first analysis, which examines the impact on the expertise values calculated by the \textit{DOE} model, we focus exclusively on the 50\% impact scenario. This choice is justified by our interest in analyzing statistical metrics, such as the overall average impact on \textit{DOE} values. 

\section{Results} \label{sec:results}

Following the study design described, we present the results in two parts. First, we provide a statistical analysis that characterizes the conversations with ChatGPT and the code integration. Second, we evaluate the impact of GenAI usage on developer expertise values, as calculated using the Degree of Expertise (DOE) model, followed by an assessment of its effect on the Truck Factors.

\subsection{Code Integration Statistics} \label{sec:code_integration_statistics}

First, to characterize the conversations in the shared links, we analyzed the number of \textit{turns} involved in the proposed solutions. A \textit{turn} consists of a user prompt followed by a ChatGPT response, following the same terminology used by Hao et. al \cite{hao2024empirical}, for example. The distribution of the number of turns in the shared solutions has a first quartile (Q1) of 4, a median (Q2) of 8, a third quartile (Q3) of 16, and a mean of 14.29.

Second, we analyzed the distribution of the percentage of copied code. The distribution has a mean of 39\%, with the first quartile (Q1) at 14\%, the median (Q2) at 31\%, and the third quartile (Q3) at 66\%. Table~\ref{tab:copy_code_language} presents a breakdown of the mean copy percentage and the number of file–shared link pairs, by programming language. As there is only one Ruby file in the dataset, we excluded it from this analysis and the table.

As described in Section~\ref{sec:impact_measurement_design}, the impact analysis requires a code copy rate to simulate the attribution of authorship to GenAI. For this purpose, in the following sections, we adopt the mean of the code copy rate distribution of 39\% as the \textit{uniform code copy rate}.

\begin{table}[h]
\centering
\caption{Percentage of copied code by programming language.}
\begin{tabular}{@{}lll@{}}
\toprule
\textbf{Language} & \textbf{Percentage} & \textbf{Num} \\ \midrule
C                 & 59\%                & 53           \\
C\#               & 51\%                & 59           \\
Shell             & 44\%                & 69           \\
Java              & 42\%                & 175          \\
Python            & 41\%                & 612          \\
C++               & 38\%                & 186          \\
JavaScript        & 34\%                & 373          \\
TypeScript        & 33\%                & 109          \\
PHP               & 32\%                & 35           \\ \bottomrule
\end{tabular}
\label{tab:copy_code_language}
\end{table}

\subsection{Impact on Degree of Expertise}

In this initial analysis, we focus exclusively on the 50\% impact scenario, as described in Section \ref{sec:impact_measurement_design}. This choice is justified by our interest in the overall average impact on \textit{DOE} values rather than individual cases explored in the Truck Factor analysis. Table \ref{tab:distribution_doe_affected_doe_diff} presents the distribution of \textit{DOE} values across developer-file pairs affected by GenAI usage, both with and without code copying. The third column (\textbf{Difference}) shows the distribution of the differences between the \textit{DOE} values in these two conditions. 

\begin{table}[h]
\centering
\caption{Quartile distribution of original DOE values, DOE values affected by code copying, and their differences.}
\begin{tabular}{@{}cccc@{}}
\toprule
\textbf{}     & \textbf{Original DOE} & \textbf{Copy Affected DOE} & \textbf{Difference} \\ \midrule
\textbf{Q1}   & 2.248                 & 2.118                      & 0.113               \\
\textbf{Q2}   & 2.793                 & 2.666                      & 0.118               \\
\textbf{Q3}   & 3.397                 & 3.277                      & 0.160               \\
\textbf{Mean} & 2.842                 & 2.716                      & 0.126               \\ \bottomrule
\end{tabular}
\label{tab:distribution_doe_affected_doe_diff}
\end{table}

The project-level analysis also did not reveal any significant differences. We calculated the mean difference in \textit{DOE} values for each project with and without copied code. The standard deviation of these means was low (0.0017), indicating that the values were tightly clustered around the overall mean of 0.125. Similarly, when grouping projects by programming language, the mean differences also showed no notable variation.

We also segmented the analysis by comparing the differences in \textit{DOE} values between core and peripheral developers. Core developers, defined here as those identified in the Truck Factor, experienced smaller losses, with an average reduction of 0.124. In contrast, peripheral developers showed a slightly higher average loss of 0.126. A \textit{Wilcoxon Signed-Rank Test} comparing the \textit{DOE} differences between these two groups yielded a statistically significant result (p < 0.005), suggesting a systematic disparity in the impact.

We also applied the \textit{Wilcoxon Signed-Rank Test} to the distribution of \textit{DOE} variations within each project, finding a statistically significant difference from zero in all cases (p < 0.005). This suggests that, despite the small magnitude of individual changes, there is a consistent and systematic effect across projects. The following sections explore how these differences affect higher-level metrics.

\subsection{Impact on Truck Factor}

First, we present a Truck Factor analysis of the selected projects. The distribution of Truck Factor values shows a mean of 108.75 and a median of 17. Due to the presence of outliers, such as \textit{torvalds/linux} and \textit{ohmyzsh/ohmyzsh}, the median offers a more representative measure of central tendency. The original (unimpacted) Truck Factor values are listed in the \textbf{TF} column of Table~\ref{tab:target_repositories_impact}.

To evaluate the impact of GenAI usage on these values, we consider two main aspects: (1) whether the Truck Factor value itself changes, and (2) whether the value remains stable but the developer ranking is affected. To quantify changes in developer rankings, we apply the Kendall Rank Correlation Coefficient, which measures the similarity between two ranked lists \cite{abdi2007kendall}.

We computed 120 Truck Factor values, covering 24 projects under 5 scenarios. The results are shown in the \textbf{TF\_\textit{N}} columns, where \textit{N} indicates the proportion of impacted files, as illustrated in Figure~\ref{fig:gen_ai_impact_analysis}. Of these, 87 values (73\%) changed, impacting 20 of the 24 projects. Among the 87 changes, 86 showed a reduction in the Truck Factor, with a median decrease of 2 developers. Only one case showed an increase, by a single developer.

\begin{table}[h]
\centering
\caption{Truck Factor values of the target repositories across different impact scenarios.}
\resizebox{\columnwidth}{!}{%
\begin{tabular}{@{}lcccccc@{}}
\toprule
\multicolumn{1}{c}{\textbf{Repository}}                                 & \textbf{TF} & \textbf{TF\_10} & \textbf{TF\_20} & \textbf{TF\_30} & \textbf{TF\_40} & \textbf{TF\_50} \\ \midrule
discourse/discourse                                                     & 23          & 22              & 22              & 21              & 21              & 21              \\
electron/electron                                                       & 16          & 15              & 15              & 15              & 15              & 15              \\
facebook/react                                                          & 6           & 6               & 7               & 6               & 6               & 6               \\
facebook/react-native                                                   & 68          & 66              & 66              & 65              & 64              & 60              \\
\begin{tabular}[c]{@{}l@{}}freeCodeCamp/\\ freeCodeCamp\end{tabular}    & 15          & 14              & 13              & 13              & 13              & 13              \\
godotengine/godot                                                       & 58          & 56              & 55              & 56              & 52              & 52              \\
huginn/huginn                                                           & 7           & 6               & 7               & 6               & 6               & 6               \\
jellyfin/jellyfin                                                       & 13          & 13              & 12              & 11              & 11              & 11              \\
laravel/framework                                                       & 212         & 189             & 188             & 189             & 181             & 173             \\
mastodon/mastodon                                                       & 5           & 5               & 5               & 4               & 4               & 4               \\
microsoft/PowerToys                                                     & 21          & 20              & 20              & 19              & 19              & 18              \\
microsoft/terminal                                                      & 6           & 6               & 6               & 6               & 6               & 6               \\
microsoft/vscode                                                        & 18          & 17              & 17              & 17              & 16              & 16              \\
netdata/netdata                                                         & 3           & 3               & 3               & 3               & 3               & 3               \\
ohmyzsh/ohmyzsh                                                         & 845         & 632             & 615             & 649             & 630             & 641             \\
\begin{tabular}[c]{@{}l@{}}PowerShell/\\ PowerShell\end{tabular}        & 13          & 13              & 12              & 12              & 12              & 12              \\
rails/rails                                                             & 455         & 399             & 400             & 391             & 380             & 374             \\
redis/redis                                                             & 33          & 31              & 31              & 30              & 30              & 30              \\
\begin{tabular}[c]{@{}l@{}}Significant-Gravitas/\\ AutoGPT\end{tabular} & 6           & 6               & 6               & 6               & 6               & 6               \\
tensorflow/tensorflow                                                   & 149         & 146             & 146             & 144             & 144             & 144             \\
torvalds/linux                                                          & 604         & 593             & 589             & 580             & 574             & 564             \\
twbs/bootstrap                                                          & 22          & 18              & 19              & 17              & 17              & 19              \\
vercel/next.js                                                          & 10          & 10              & 10              & 10              & 10              & 9               \\
vuejs/vue                                                               & 2           & 2               & 2               & 2               & 2               & 2               \\ \bottomrule
\end{tabular}%
}
\label{tab:target_repositories_impact}
\end{table}

Additionally, 85 Truck Factors (71\%) exhibited differences in their ranking order across 21 projects. The distribution of $\tau$ (tau) values, representing ranking similarity, has a mean of 0.43, with the first quartile (Q1) at 0.26, the median (Q2) at 0.37, and the third quartile (Q3) at 0.60. Among the 21 affected projects, 12 began to show differences, either in the value of the Truck Factor or in the order, in the first 10\% impact scenario. Seven projects exhibited changes specifically at the 20\% impact scenario, while one project showed changes only at the 30\% scenario, and another at the 50\% scenario.

We found a moderate positive correlation of $\rho = 0.41$ between the original Truck Factor size and the frequency of changes across scenarios. However, this does not mean that changes were exclusive to projects with higher Truck Factors. To investigate whether projects with lower Truck Factors were also affected, we selected those with a Truck Factor less than or equal to 7, approximately the first quartile (6.75) of the original Truck Factor distribution. This subset represents 35 (29\%) of the 120 calculated values. Even within this group, 17 (49\%) exhibited changes, either in their ranking order or in their Truck Factor value.

\section{Discussion} \label{sec:discussion}

Firstly, regarding the number of turns in the mined conversations, our results differ from those reported by Hao et al., where most conversations consisted of only a single turn \cite{hao2024empirical}. A key difference lies in the datasets: our study focuses specifically on code generation that was at least partially integrated into real projects. A more comparable reference is the work by Jin et al., who reported an average of 10.4 turns per code generation conversation \cite{jin2024can}. In contrast, our dataset shows a considerably higher average of 14.291 turns. However, the 100th percentile of our distribution is 326, indicating the presence of outliers. In this context, the median value of 8 is a more representative central tendency and aligns more closely with results from other studies.

The data on the percentage of copied code by programming language suggests that lower-level languages, such as \textit{C} and \textit{Shell}, exhibit the highest rates. In contrast, higher-level languages like \textit{JavaScript} and \textit{TypeScript} show lower percentages. This trend may be associated with the level of abstraction provided by each language, with higher-level languages tending to be more concise, potentially reducing the need for directly copying code snippets.

Regarding the impact analysis, the difference in developers' Degree of Expertise values within the files they contributed to was consistent but small, as shown in Section \ref{sec:results}. This outcome is expected, given that the simulation removed 39\% of only one variable in the \textit{DOE} model, which is not the most significant for assessing knowledge, as discussed in Section \ref{sec:background}. Among the stratifications conducted, none revealed a notable difference, except that core developers appeared to lose slightly less knowledge and were less affected overall. However, even these small changes were reflected in the Truck Factor calculations, impacting both high and low values, with a strong tendency toward decreasing the Truck Factor. In such scenarios, the algorithm tends to suggest a greater concentration of knowledge among key developers.

This finding demonstrates that the Truck Factor is sensitive to changes in developers' expertise, \textit{raising concerns about its reliability when GenAI-generated code is involved}. However, this issue extends beyond the Truck Factor, as authorship metrics are widely employed in various software engineering tasks. For instance, measures such as \textit{commits} and \textit{lines of code} are used to assess technical expertise for recruitment \cite{atzberger2022codecv}, to identify appropriate code reviewers \cite{hannebauer2016automatically}, and even as indicators of developer productivity \cite{beller2025s}. We anticipate that all of these applications may be impacted as the integration of GenAI tools and automated code generation becomes more prevalent in software development.

\section{Conclusion and Future Work} \label{sec:conclusion}

In this paper, we presented a study on the impact of using Generative Artificial Intelligence (GenAI) for code generation in a developer expertise identification model and a Truck Factor algorithm. Based on statistics regarding the percentage of code copied from ChatGPT, our simulations suggest that expertise identification models are sensitive to attribution loss. Although the quantitative reduction in expertise values is relatively small, applications such as the Truck Factor algorithm are still affected, even under scenarios of limited GenAI usage. These results highlight a potential lack of confidence in such metrics as GenAI tools become more prevalent in software development.

This is an exploratory study that opens several avenues for future research. We plan to investigate other knowledge models and Truck Factor algorithms using a similar methodology. Furthermore, we aim to explore developers' perceptions of this issue. To that end, we intend to survey to understand how developers integrate GenAI-generated code into their projects and how they perceive its impact on source code knowledge and expertise attribution.

\section{Artifact Availability}

We have made all artifacts from our study publicly available at \url{https://doi.org/10.5281/zenodo.15334705} \cite{cury_2025_15334705}.

\bibliographystyle{ACM-Reference-Format}
\bibliography{sample-base}

\end{document}